\documentclass[aps,pra,twocolumn,superscriptaddress]{revtex4-1}
\usepackage{graphicx}
\usepackage{hyperref}
\usepackage{amsmath}
\usepackage{amssymb}
\usepackage{epstopdf}
\usepackage{multirow}
\usepackage{threeparttable}
\usepackage{xcolor}
\usepackage{ulem}

\begin{document}

\title{High-order mid-infrared nonlinear topological differentiator}
\author{Jixi Zhang}
\affiliation{State Key Laboratory of Precision Spectroscopy, and Hainan Institute, East China Normal University, Shanghai 200062, China}

\author{Kun Huang}
\email{khuang@lps.ecnu.edu.cn}
\affiliation{State Key Laboratory of Precision Spectroscopy, and Hainan Institute, East China Normal University, Shanghai 200062, China}
\affiliation{Chongqing Key Laboratory of Precision Optics, Chongqing Institute of East China Normal University, Chongqing 401121, China}
\affiliation{Collaborative Innovation Center of Extreme Optics, Shanxi University, Taiyuan, Shanxi 030006, China}

\author{Shina Liao}
\affiliation{State Key Laboratory of Precision Spectroscopy, and Hainan Institute, East China Normal University, Shanghai 200062, China}

\author{Zhuohang Wei}
\affiliation{State Key Laboratory of Precision Spectroscopy, and Hainan Institute, East China Normal University, Shanghai 200062, China}

\author{Jianan Fang}
\affiliation{State Key Laboratory of Precision Spectroscopy, and Hainan Institute, East China Normal University, Shanghai 200062, China}
\affiliation{Chongqing Key Laboratory of Precision Optics, Chongqing Institute of East China Normal University, Chongqing 401121, China}

\author{Heping Zeng}
\email{hpzeng@phy.ecnu.edu.cn}
\affiliation{State Key Laboratory of Precision Spectroscopy, and Hainan Institute, East China Normal University, Shanghai 200062, China}
\affiliation{Chongqing Key Laboratory of Precision Optics, Chongqing Institute of East China Normal University, Chongqing 401121, China}
\affiliation{Shanghai Research Center for Quantum Sciences, Shanghai 201315, China}
\affiliation{Chongqing Institute for Brain and Intelligence, Guangyang Bay Laboratory, Chongqing, 400064, China}

\begin{abstract}
High-order edge-enhanced imaging enables precise feature localization and effective background suppression, offering a powerful tool for real-time recognition and high-contrast visualization. Extending this capability to the mid-infrared (MIR) regime is particularly valuable for applications such as biomedical diagnostics, material inspection, and remote sensing, yet remains limited by inadequate spatial-frequency modulation fidelity and low detection sensitivity. Here, we demonstrate a high-sensitivity MIR upconversion differentiator operating at 3 $\mu$m, which achieves isotropic high-order edge enhancement by optically imprinting topological complex-amplitude patterns onto MIR Fourier components via nonlinear parametric interaction. Vortex transfer functions $t(k_r, \phi) \propto k_r^\ell e^{i\ell\phi}$ are precisely encoded on a phase-only spatial light modulator to enable tunable MIR differentiation from first- to fourth- order, with real-time switching at up to 60 Hz. Benefiting from a low-noise upconversion process and a single-photon-sensitive silicon camera, the system achieves high-contrast edge imaging under low-light conditions. Experimental results confirm accurate edge extraction and background suppression for both amplitude and phase objects, hence underscoring its potential for noninvasive diagnostics and label-free material analysis.
\end{abstract}

\maketitle

\section{Introduction}
Edge detection is a fundamental operation in image processing, essential for extracting structural features such as object contours and boundaries \cite{Zhou2020NP}. While conventional digital methods rely on post-acquisition convolution and intensive computation, optical differentiators perform spatial frequency manipulation intrinsically during light propagation, enabling real-time edge detection with ultralow latency and minimal power consumption \cite{Solli2015NP}. Moreover, the field-resolved operation of optical analog differentiation enables access to phase and polarization information, which is advantageous for visualizing transparent or anisotropic specimens \cite{Zhou2020SA, Wang2024NREE}. With their ultrafast parallel response and multidimensional analytical capability, optical differentiators are increasingly recognized as powerful tools in biomedical imaging, computer vision, and nondestructive inspection \cite{He2022Nano}.

Beyond first-order edge enhancement, high-order optical differentiation offers improved performance by suppressing mid- and low-frequency backgrounds and highlighting curvature and fine topological features \cite{Liu2022NC, Huo2024NC}. Existing implementations typically rely on complex amplitude filtering in 4f system \cite{Chen2024OE, Chen2025LPR} or nanophotonic structures exploiting physical mechanisms such as plasmonic resonance, spin-orbit coupling, and the photonic Spin Hall effect \cite{Zhu2021NC, Qiu2025NC, Deng2024NC, Cotrufo2024NC}. With the advancement of metasurfaces and integrated photonics, these platforms enable programmable and compact high-order analog operations \cite{He2022Nano}. To date, the development of all-optical high-order spatial differentiation techniques has primarily focused on the visible and near-infrared (NIR) spectral regions, benefiting from the exceptional optical response of nanophotonic materials and well-established optical manipulation technologies available at these wavelengths. 

In particular, the mid-infrared (MIR) spectral region (2-12 $\mu$m) hosts abundant molecular vibrational resonances and coincides with multiple atmospheric transmission windows, positioning it as a key regime for label-free chemical sensing, biomedical diagnostics, and thermal remote imaging \cite{Vodopyanov2020Book}. Extending high-order optical edge detection into this band could thus unlock powerful novel possibilities for contrast-enhanced, information-rich infrared imaging. However, this promise remains largely unrealized due to several fundamental challenges. First, 
existing nanophotonic materials are plagued by the limited optical tunability and weak MIR response \cite{Herrmann2020JAP}, severely constraining the spatial-frequency modulation fidelity required for high-order differentiation. Second, thermal background radiation and the performance limitations of MIR detectors make high-sensitivity, low-noise detection difficult, especially under ambient conditions \cite{Wang2019Small}. Furthermore, long-wavelength diffraction effects and the lack of broadband, reconfigurable MIR-compatible phase filters impose additional constraints on resolution, flexibility, and system scalability \cite{Edgar2019NP}.

Recently, frequency upconversion has emerged as  a powerful approach for MIR imaging by converting MIR signals into visible or NIR wavelengths in a nonlinear crystal, which enables fast and sensitive detection with high-performance silicon photodetectors \cite{Barh2019AOP, Zeng2023LPR, Liu2022Optica, Huang2022NC, Dam2012NP}. In 4f-configured systems, the pump's spatial structure is directly mapped onto the MIR spatial-frequency domain through three-wave mixing, allowing precise optical modulation in the Fourier plane \cite{Huang2022NC, Dam2012NP, Zheng2024Optica, Wang2023NC, Junaid2020AO}. Specifically, by projecting a first-order spiral phase mask onto the Fourier plane, a Hilbert transform is applied to the signal field, enabling upconversion edge-enhanced imaging \cite{Wang2021LPR, Qiu2018Optica, Liu2019PRAppl, Gao2025OE, Li2024OE}. However, pure spiral phase filtering inherently restricts differentiation to the first order, yielding broad edges and limited localization accuracy. Simply increasing the topological charge $\ell$ amplifies angular contrast but does not realize higher-order differentiation \cite{Li2021OE}. On the other hand, pump profiles with purely radial amplitude modulation, such as $t(k_r) \propto k_r^{\ell}$, act as Laplacian-like filters that provide isotropic enhancement only for even orders \cite{Chen2024OE}. These approaches lack generality and tunability across arbitrary differentiation orders. Consequently, isotropic high-order edge enhancement in the MIR remains an open and pressing challenge.

In this work, we report and experimentally demonstrate, for the first time to our knowledge, a tunable MIR upconversion differentiator capable of performing arbitrary isotropic high-order spatial differentiation. The core innovation lies in encoding a complex modulation profile of the form $t(k_r, \phi) \propto k_r^\ell e^{i\ell\phi}$ onto a phase-only spatial light modulator (SLM), which transforms a Gaussian pump into a topological differentiator. Comparative results show that only this complex amplitude design achieves isotropic and reconfigurable high-order differentiation. Simulations and experiments demonstrate high-fidelity differentiation up to the fourth order, with strong agreement between theoretical predictions and measured results. Leveraging the programmable flexibility of the SLM, the differentiation order can be dynamically switched in real time without mechanical adjustments. Furthermore, edge enhancement is validated for both amplitude and phase objects, demonstrating strong background suppression and precise recovery of fine features. These results highlight the potential of this approach for contrast-enhanced and non-invasive analysis in chemical and biomedical imaging.

\begin{figure*}[t!]
	\centering
	\includegraphics[width=0.90\textwidth]{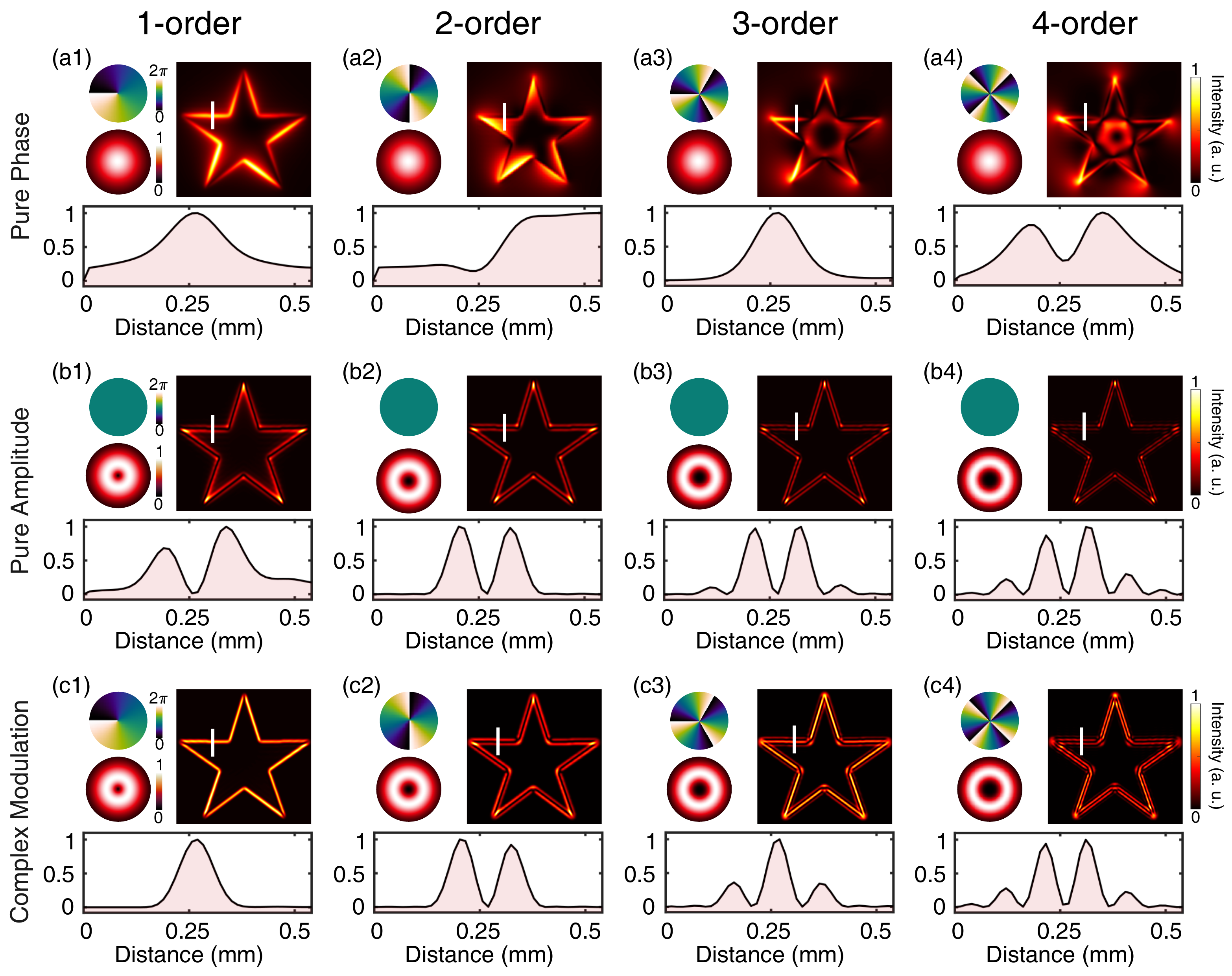}
	\caption{Nonlinear upconversion spatial differentiation under different pump filter configurations. Each image consists of three elements: the complex amplitude of the pump in the Fourier plane, the corresponding upconverted image intensity, and the edge intensity profile extracted along the white solid line. (a1-a4) Pump carrying a pure spiral phase $e^{i\ell\phi}$ with topological charge $\ell$. Only $\ell = 1$ produces isotropic edge enhancement via a radial Hilbert transform, while higher-order spiral phases primarily enhance angular contours. (b1-b4) Pump with amplitude shaped as $t(k_r) \propto k_r^\ell$ and a flat phase, acting as a radial differentiator. Isotropic enhancement occurs only at even orders. (c1-c4) Pump with complex amplitude $k_r^\ell e^{i\ell\phi}$, forming a topological differentiator that yields isotropic edge enhancement at all orders. The number of peaks in the edge profile matches the differentiation order.}
	\label{fig1}
\end{figure*}

\section{Basic principle}
The core principle of the nonlinear spatial differentiator lies in sum-frequency generation (SFG), where the angular frequencies of the interacting fields satisfy $\omega_s + \omega_p = \omega_u$. Under phase-matching conditions, the upconverted field is proportional to the product of the signal and pump fields. The nonlinear wave mixing enables both wavelength conversion and coherent transfer of the pump's spatial information to the MIR signal. This intrinsic property allows the pump to act as a complex spatial filter in the Fourier domain. While prior works have demonstrated MIR edge enhancement using pure phase (e.g., spiral-phase) \cite{Wang2021LPR, Liu2019PRAppl, Qiu2018Optica} or amplitude-only modulation schemes \cite{Junaid2020AO}, such approaches are limited to first-order differentiation or specific symmetry constraints \cite{Li2021OE}. Here, we extend the concept to complex-field pump engineering, simultaneously modulating amplitude and phase to construct high-order isotropic differentiators. This nonlinear optical filtering mechanism enables high-fidelity edge enhancement with strong background suppression and high detection sensitivity, which is essential for robust MIR imaging of weak amplitude and phase objects.

In a 4f upconversion imaging system, the electric fields of the interacting beams are related as follows \cite{Barh2019AOP}
\begin{align}
  E_u(x,y) &\propto \mathcal{F}^{-1}\!\left[\,\tilde{E_s}(k_x,k_y)\,\tilde{E_p}(k_x,k_y)\,\right] \label{eq2} \\
  &= \iint \tilde{E_s}(k_x,k_y)\,\tilde{E_p}(k_x,k_y)\,e^{i(k_x x + k_y y)}\,\mathrm{d}k_x\,\mathrm{d}k_y \nonumber \ ,
\end{align}
where $\mathcal{F}^{-1}$ denotes the inverse Fourier transform, $\tilde{E}_s(k_x,k_y)$ is the spatial-frequency spectrum of the MIR signal. $\tilde{E}_p(k_x,k_y)$ denotes the pump field in $k$-space, obtained by mapping its transverse spatial distribution in the Fourier plane through the coordinate transformation $k_r = \tfrac{2\pi}{\lambda_s f} r$, where $f$ is the focal length of the Fourier lens and $\lambda_s$ is the signal wavelength. Hence, the pump's real-space field profile directly defines a spectral filter acting on the MIR signal, and by tailoring its complex amplitude, arbitrary spatial-frequency operations  can be implemented, which effectively turns the pump into a programmable all-optical convolution kernel for versatile nonlinear imaging tasks \cite{Fu2022LSA}. 

Specifically, the pump's spatial-frequency transfer function can be engineered as $\tilde{E}_p(k_x,k_y)=k_x+i k_y = k_r e^{i\phi}$, where $k_r=\sqrt{k_x^2+k_y^2}$ is the radial spatial frequency and $\phi=\mathrm{arctan}(k_y/k_x)$ is the azimuthal angle. Under such design, the upconverted field corresponds to the complex gradient of the input signal
\begin{equation}
E_{{u}}(x,y) \propto \frac{\partial E_s}{\partial x} + i\,\frac{\partial E_s}{\partial y} \ ,
   \label{eq3}
\end{equation}
which represents isotropic first-order spatial differentiation. Accordingly, the upconverted intensity is given by
\begin{equation}
I_{{u}}(x,y) 
  = \big|E_{u}(x,y)\big|^{2}
   \propto \left|\frac{\partial E_s}{\partial x}\right|^{2}
    + \left|\frac{\partial E_s}{\partial y}\right|^{2} \ ,
	 \label{eq4}
\end{equation}
showing that the output captures the squared gradient magnitude of the input field. Since the filter's amplitude depends only on $k_r$ and its phase varies linearly with $\phi$, the resulting transfer function exhibits cylindrical symmetry. This ensures uniform first-order edge enhancement across all orientations. A complete mathematical derivation is provided in Supplementary Note 1.
 
Furthermore, when the complex amplitude of the pump filter is designed as $\tilde{E}_p(k_x,k_y)\propto k_r^{\ell} e^{i {\ell}\phi}$, isotropic high-order edge-enhanced imaging can be achieved with arbitrary differentiation order $\ell$. The corresponding upconverted intensity takes the form \cite{Li2022OL}
\begin{equation}
    I_u(x,y)\propto
    \Bigl|
      \bigl(\partial_x+i\partial_y\bigr)^{\ell}
      E_s(x,y)
    \Bigr|^{2} \ ,
    \label{eq5}
\end{equation}
representing the squared magnitude of the complex directional derivative of order $\ell$.

In the 4f imaging configuration, the spatial frequency $k_r$ relates to the physical radial coordinate in the Fourier plane. Hence, the required spatial-domain transfer function becomes $T(r, \phi) \propto r^\ell e^{i\ell\phi}$. Notably, Laguerre--Gaussian (LG) vortex beams inherently exhibit such a structure, featuring a radial amplitude term and a helical phase $e^{i\ell\phi}$, modulated by a Gaussian envelope \cite{Singh2024AdvPhoton}. By neglecting the generalized Laguerre polynomial, we therefore design an LG$_0^{\ell}$-like mode as follows
\begin{equation}
E_p(r, \phi) =
\left(\frac{\sqrt{2} r}{w_0}\right)^{|\ell|}
\exp \left(-\frac{r^2}{w_0^2}\right)
\exp(i\ell\phi) \ ,
\label{eq6}
\end{equation}
where $w_0$ is the beam waist. Near the beam center ($r \ll w_0$), the amplitude approximates the ideal radial profile $r^\ell$, making it well-suited for high-order spatial differentiation \cite{Chen2025LPR}. We thus employ a phase-only SLM to holographically reshape a Gaussian pump beam into an LG$_0^{\ell}$-like mode, effectively implementing a tunable topological differentiator within the upconversion imaging system.

\begin{figure*}[t!]
	\centering
	\includegraphics[width=0.9\textwidth]{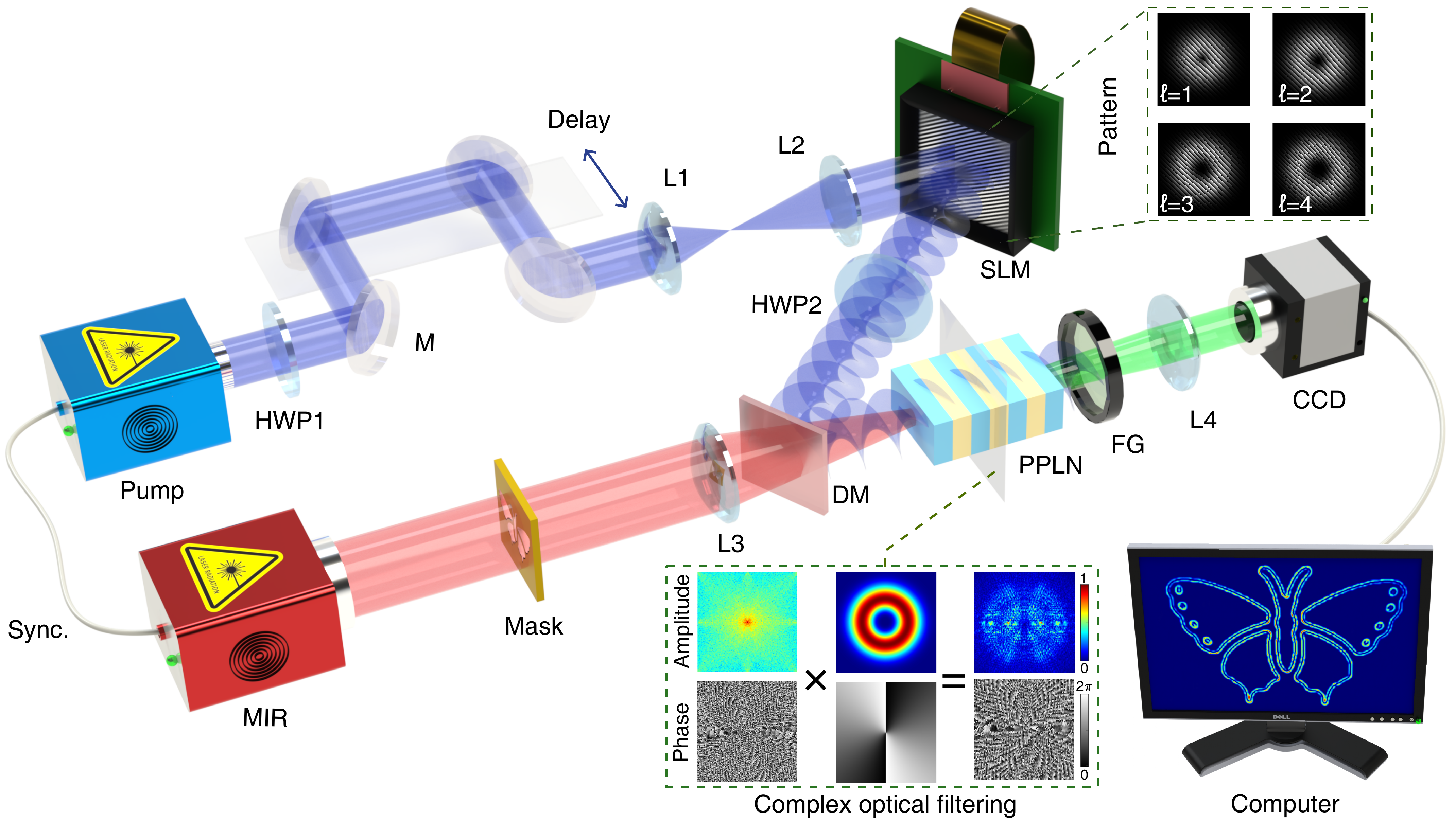}
	\caption{Experimental setup of the high-order MIR upconversion differentiation imaging system. The system employs two synchronized pulsed lasers at 1030 nm and 3070 nm as the pump and signal sources, respectively. A collimated MIR beam illuminates the object and propagates through a 4f imaging configuration formed by lenses L3 and L4. Meanwhile, a structured pump beam carrying the desired vortex complex profile is generated by a phase-only spatial light modulator (SLM). The top-right inset shows the encoded holographic phase patterns for multiple differentiation orders. The shaped vortex pump is then directed into a periodically poled lithium niobate (PPLN) crystal to perform sum-frequency generation (SFG). This nonlinear interaction enables direct mapping of the pump's complex amplitude onto the MIR spatial-frequency components at the Fourier plane, as illustrated in the bottom-right dashed box. The SFG process thus performs high-fidelity nonlinear spatial filtering for edge enhancement, while simultaneously converting the MIR signal into the visible band for sensitive detection. After spectral filtering, the upconverted edge-enhanced image is captured by a silicon-based CCD camera. HWP, half-wave plate; M, mirror; DM, dichroic mirror; L, lens; FG, filtering group; CCD, charge-coupled device.}
	\label{fig2}
\end{figure*}

To evaluate the performance of our complex pump-filter design in MIR high-order differential imaging, we conducted a comparative analysis across first- to fourth-order edge enhancement using three modulation strategies: spiral phase only, radial amplitude only, and full complex-field modulation, as summarized in Fig. \ref{fig1}. Each subpanel comprises three elements: the upper-left inset shows the pump's complex amplitude at the Fourier plane of the 4f upconversion system, the upper-right panel presents the corresponding edge-enhanced output, and the bottom plot displays the edge intensity profile along the white line. In Figs. \ref{fig1}(a1-a4), the pump carries a pure spiral phase $e^{i\ell\phi}$. For $\ell = 1$, the modulation imposes a radial Hilbert transform, yielding isotropic first-order edge enhancement. However, for $\ell \geq 2$, the response becomes increasingly angular, primarily emphasizing curvature variations and resulting in anisotropic edge outlines. Figures \ref{fig1}(b1-b4) use a pump with radial amplitude $t(k_r) \propto k_r^{\ell}$ and uniform phase, mimicking a Laplacian-type operator. While this configuration produces isotropic enhancement at even orders, odd-order cases exhibit significant response deficiencies, limiting general applicability. In contrast, Figs. \ref{fig1}(c1-c4) employ a pump with complex amplitude $k_r^{\ell}e^{i\ell\phi}$, forming a topological differentiator. This configuration achieves isotropic and order-accurate edge enhancement across all tested orders. The number of peaks observed in the edge-intensity profiles directly corresponds to the differentiation order, confirming precise spectral shaping and high-fidelity spatial-frequency control.

\section{Experimental section}
The experimental setup for the high-order MIR upconversion edge-enhanced imaging system is illustrated in Fig. \ref{fig2}. The signal and pump light sources originate from two temporally synchronized MIR and NIR lasers, which operate at a repetition rate of 14.6 MHz. The MIR beam at 3070 nm is generated via difference-frequency generation under the pulsed pumping at 1030 nm. The passive synchronization allows us to obtain dual-color sources with compact layout, simplified operation, and enhanced robustness, thus ensuring a long-term stable operation. The pulse durations for the MIR signal and NIR pump are measured to be 26 and 32 ps,  respectively. The slightly longer pump pulse is intentionally used to improve energy utilization efficiency and nonlinear conversion efficiency in the subsequent conversion process. More details about the laser system can be found in our previous work \cite{Wang2021LPR}.

Then, the generated synchronized pulses are adopted to implement MIR nonlinear differentiator based on the  coincidence-pumping configuration \cite{Huang2022NC}. The object image is formed by a transmission mask, and enters a 4f-based upconversion imaging system. A periodically poled lithium niobate (PPLN) crystal is placed at the Fourier plane, where the spatial-frequency components are optically modulated by the pump field. The nonlinear crystal has geometric dimensions of 10$\times$12$\times$2 mm$^3$ (length$\times$width$\times$thickness) and is fabricated with four poling periods, among which 20.9~$\mu$m is chosen in the experiment. It is mounted in an oven stabilized at 48.6~$^\circ$\text{C} to approach the phase-matching condition. The nonlinear conversion efficiency is measured to be about 5\% at a pump power of 700 mW.

To facilitate the nonlinear differentiation operation, the pump beam is steered into a phase-only SLM (HOLOEYE, PLUTO-2.1). The SLM is equipped with 1920$\times$1080 pixels with a pitch size of 8 $\mu$m, and can support a frame refreshing rate up to 60 Hz.  The programmable SLM enables precise and agile modulation of the optical field, which facilitates the implementation of flexible and high-fidelity imaging processing functionalities. Specifically, a Gaussian pump beam is reshaped into LG$_0^{\ell}$-like modes via computer-generated holography (CGH) loaded onto an SLM. To convert phase modulation into complex-amplitude shaping, a blazed grating carrying a vortex envelope is superimposed with the spiral phase. The amplitude information can be indirectly encoded by exploiting the monotonic relationship between the phase depth of the blazed grating and the diffraction efficiency of its first diffraction order. The desired structured pump field was obtained in the first diffraction order and directed into the nonlinear crystal through a short propagation path to suppress vortex beam divergence. Detailed procedures for generating LG$_0^{\ell}$-like  modes are provided in Supplementary Note 2.

Finally, the pump path is finely adjusted to optimize the temporal overlap with the MIR pulse, producing sum-frequency radiation centered at 771 nm. The upconverted signal passes through a series of spectral filters to eliminate the pump-induced parametric fluorescence and environmental background noises. A silicon-based CCD camera is used to capture the upconversion image, which offers superior performances over the MIR focal plane arrays, with desirable features of high detection efficiency, fast frame rate, and room-temperature operation. 
 
\begin{figure}[t!]
	\includegraphics[width=1 \columnwidth]{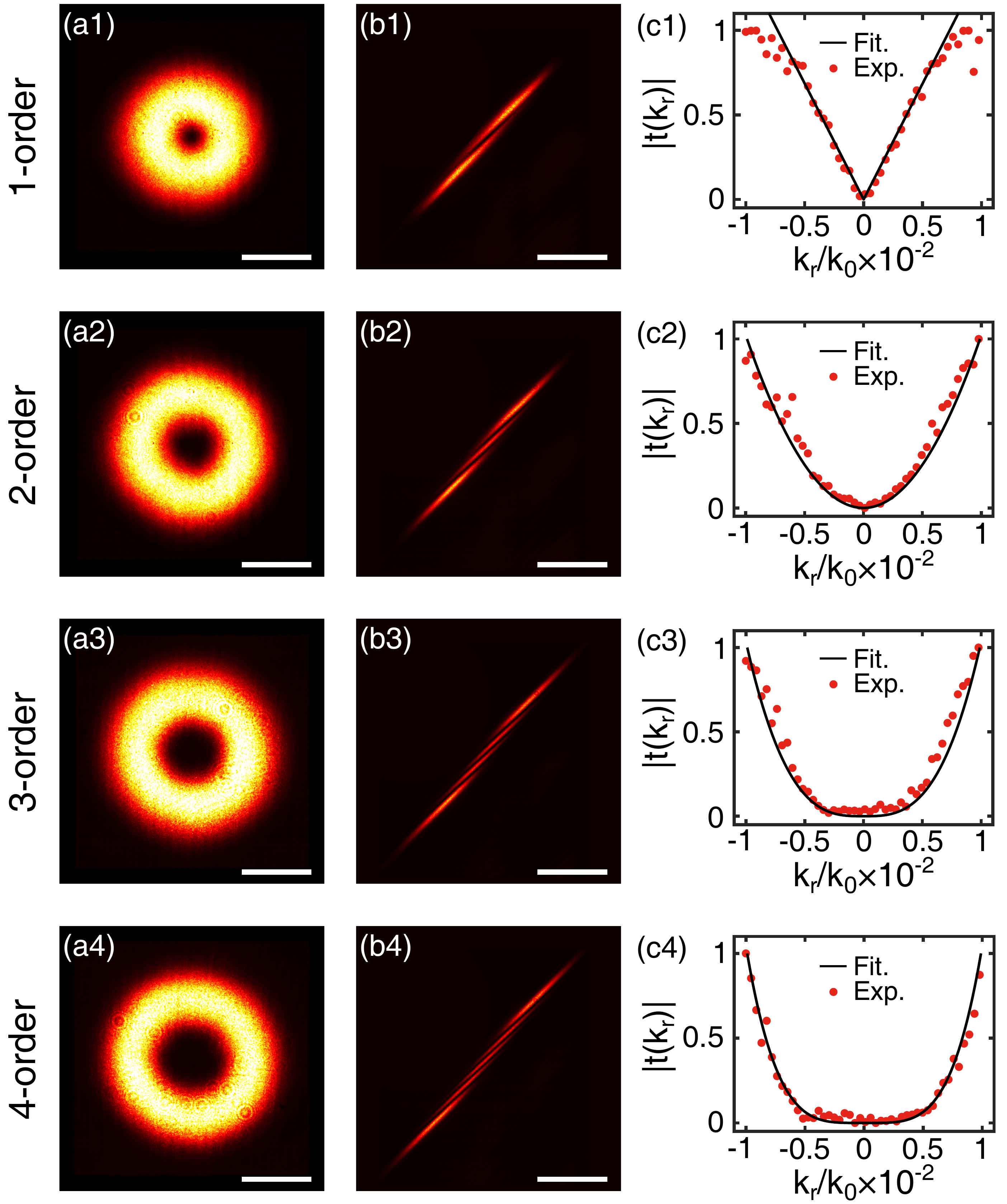}
	\caption{Verification of multiple-order nonlinear topological differential kernels. (a1-a4) Measured intensity distributions of first- to fourth-order vortex beams in the Fourier plane of the upconversion imaging system. Scale bar: 1 mm. (b1-b4) Corresponding intensity patterns after being relayed onto a secondary SLM encoded with a  saddle-shaped phase. Each beam splits into $\ell + 1$ lobes, verifying the intended topological charge $\ell$. Scale bar: 1 mm. (c1-c4) Normalized radial transmittance curves $|t(k_r)|$ exhibit a power-law scaling $t(k_r)\propto k_r^\ell$, consistent with the designed coherent transfer functions as fitted by the corresponding order polynomial.}
	\label{fig3}
\end{figure}

\begin{figure*}[t!]
	\centering
	\includegraphics[width=0.73 \textwidth]{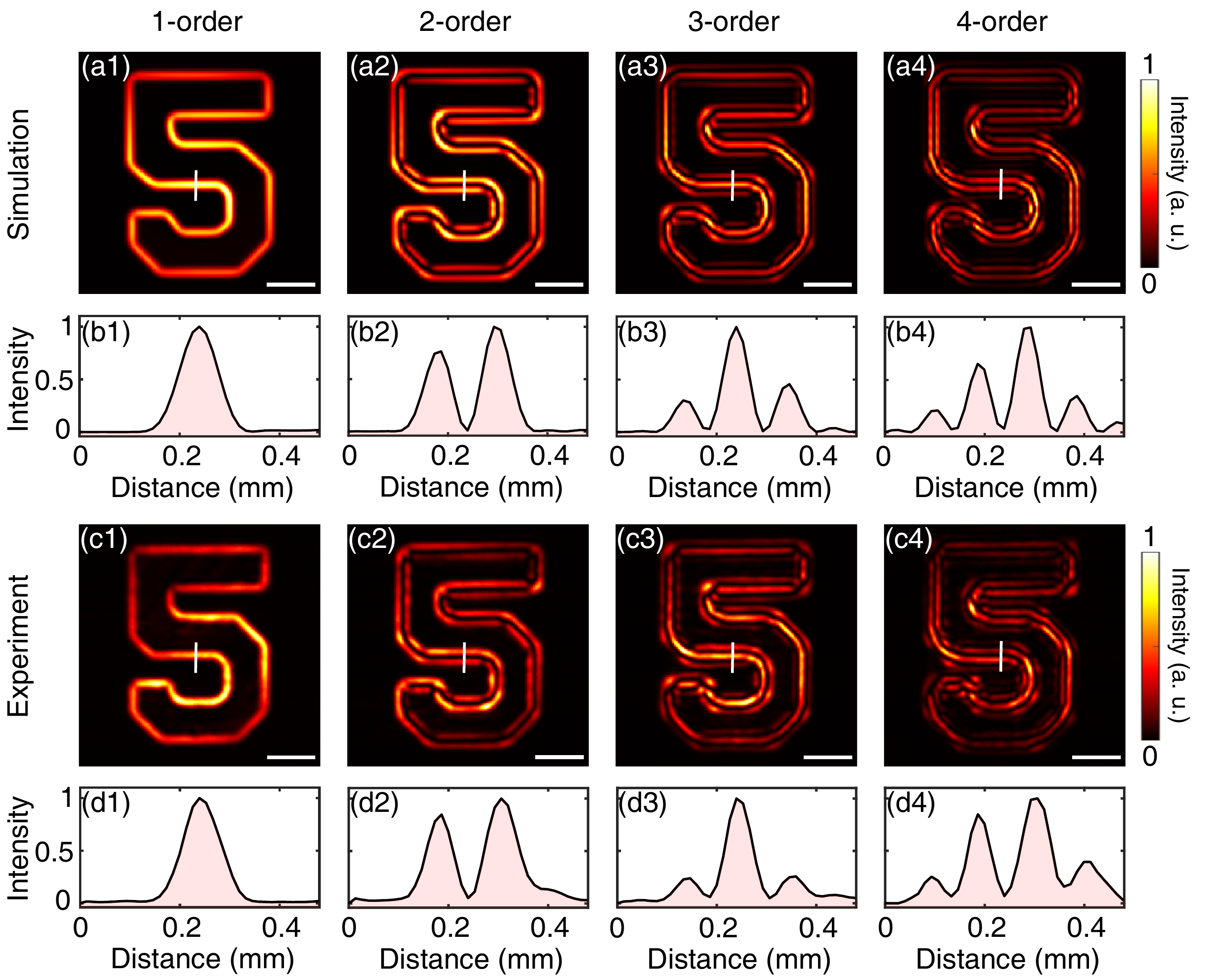}
	\caption{Multiple-order optical analog differentiation operation with nonlinear vortex filtering. (a1-a4) Simulated first- to fourth-order edge-enhanced images under experimental conditions. (b1-b4) Corresponding intensity cross-section profiles extracted along the white lines in (a1-a4), respectively. For each order, the number of intensity peaks equals the differentiation order, confirming isotropic high-order edge enhancement. (c1-c4) Experimentally acquired edge-enhanced images under the same differentiation orders. (d1-d4) Measured intensity profiles corresponding to (c1-c4). The close agreement between experiment and simulation validates the high fidelity of the nonlinear spatial differentiation operations. The scale bar in each image corresponds to 0.5 mm.}
\label{fig4}
\end{figure*}

\section{Results and discussion}
\subsection{Characterization of multiple-order topological differential kernels}
To enable high-order, high-fidelity spatial differentiation, it is essential that the pump field in the Fourier plane precisely satisfies both the amplitude and phase modulation requirements of the desired topological filter. We therefore begin by experimentally characterizing its complex field profile to verify that the generated pump beams possess the intended spatial-frequency response. As shown in the first column of Fig. \ref{fig3}, we recorded the intensity profiles of first- to fourth-order vortex pumps using an InGaAs camera. With increasing topological charge $\ell$, the central dark core expands and the main annular energy band shifts outward, indicating stronger weighting of higher spatial-frequency components \cite{Zhu2021NC}. This behavior is consistent with enhanced edge contrast and fine-feature sensitivity, while also imposing stricter demands on imaging sensitivity due to the suppression of low-frequency content.

Then, we employed a vortex mode decomposition technique to confirm the presence of the intended spiral phase and verify the topological charge. Specifically, a second SLM was inserted to encode a saddle-shaped phase onto the pump beam. This special phase is mathematically equivalent to two orthogonal cylindrical lenses, imposing an anisotropic modulation on the vortex beam and breaking its rotational symmetry \cite{Yang2022APN}. Upon subsequent free-space diffraction, this diagnostic modulation induces the vortex beam to split into a number of bright lobes, where the lobe count serves as a direct signature of the vortex charge. As shown in the second column of Fig. \ref{fig3}, we observe $\ell+1$ distinct lobes for each order, confirming the accurate generation of the desired topological charge. A detailed description of this topological charge determination technique, along with numerical simulations, is provided in Supplementary Note 3. 

We further quantified the pump's amplitude transfer function by extracting the square root of the recorded intensity, yielding the radial amplitude profile. Noting that the spatial coordinate at the Fourier plane is linearly proportional to the transverse momentum ($k$-space) of the pump beam, we rescaled the horizontal axis to the dimensionless spatial frequency $k_r/k_0$, where $k_0 = 2\pi/\lambda_s$. This normalization facilitates a clearer interpretation of the filter's spatial-frequency response. The experimental data (red dots) and power-law fits (black curves) demonstrate good agreement in the low-frequency region, confirming the transfer function follows the target form $t(k_r) \propto (k_r/k_0)^\ell$. Together with the prior phase verification, these results validate that the engineered vortex pump fields meet the design criteria for implementing multi-order topological differentiation in MIR upconversion imaging.
\newline

\subsection{Performance of multiple-order MIR nonlinear differential imaging}
We now turn to evaluate the performance of the MIR upconversion system in implementing high-order spatial differentiation. As the test object, the digit ``5" from the first group of a 1951 USAF resolution target was used to provide well-defined amplitude contrast. The mask dimensions are approximately 2.5 $\times$1.8 mm$^2$. Figures \ref{fig4}(a1-a4) show numerically simulated differential images from first- to fourth-order, while Figs. \ref{fig4}(c1-c4) display the corresponding experimental results. The intensity cross-sections along the white guide lines are plotted in Figs. \ref{fig4}(b1-b4) and (d1-d4), for simulations and experiments respectively. Across all orders, the experimental data closely match the simulations, demonstrating the anticipated edge-enhancement characteristics. Specifically, increasing the kernel order leads to a progressive transformation: the image evolves from a single bright outline to multiple finer edge contours, with enhanced suppression of low-frequency interior content and improved edge contrast. The cross-sections reveal a one-to-one correspondence between the number of intensity peaks and the order of differentiation.

Among the results, the second-order image shows a double-peak pattern that symmetrically brackets the object boundary, enabling improved edge localization compared to first-order differentiation. Higher-order kernels (third and fourth) accentuate curvature and turning points, though at the cost of more prominent ringing artifacts and increased sensitivity to noise. Importantly, the upconversion imaging system benefits from ultrashort optical gating provided by the ultrafast pump pulses, along with a high-rejection spectral filtering chain and a low-noise silicon camera with single-photon sensitivity \cite{Wang2021LPR}. These factors collectively ensure sufficient detection sensitivity for real-time fourth-order edge imaging, with an integration time as short as 10 ms per frame. In the experiment, the MIR power densities for first- and fourth-order operations are set at 0.1 mW/cm$^2$ and 1 mW/cm$^2$, respectively. Notably, the reconfigurable nature of the system allows rapid switching between differentiation orders. Once optical alignment is stabilized, the kernel order can be changed simply by updating the holographic pattern displayed on the SLM. The switching rate is ultimately limited by the SLM's refresh speed, allowing up to 60 Hz real-time operation.

In the ideal case, the multi-peak profiles from high-order differentiation would exhibit perfect symmetry, with zero background between adjacent peaks. In practice, however, slight asymmetries appear in both simulations and experiments. These arise due to deviations from the ideal 4f configuration: the finite size and angular acceptance of the nonlinear crystal impose an anisotropic truncation on the spatial-frequency spectrum, which distorts the effective point spread function and introduces lateral asymmetry. Consequently, the peak positions shift slightly, and their intensities become unequal. A detailed analysis of this asymmetry is provided in Supplementary Note 4.

 In principle, higher-order ($\ell>4$) differentiation can be achieved by designing a topological pump differentiator with the corresponding order, but this also introduces greater challenges to the sensitivity of the imaging system. Furthermore, by adjusting the poling period or tuning the operating temperature of the PPLN crystal, the system can support broadband MIR imaging. Moreover, employing nonlinear materials such as orientation-patterned gallium arsenide (OP-GaAs) or barium gallium selenide (BGSe) can further extend the operational window into the far-infrared region \cite{Demur2017OL, Liu2022Optica}.

\begin{figure}[t!]
	\centering
	\includegraphics[width=0.95\columnwidth]{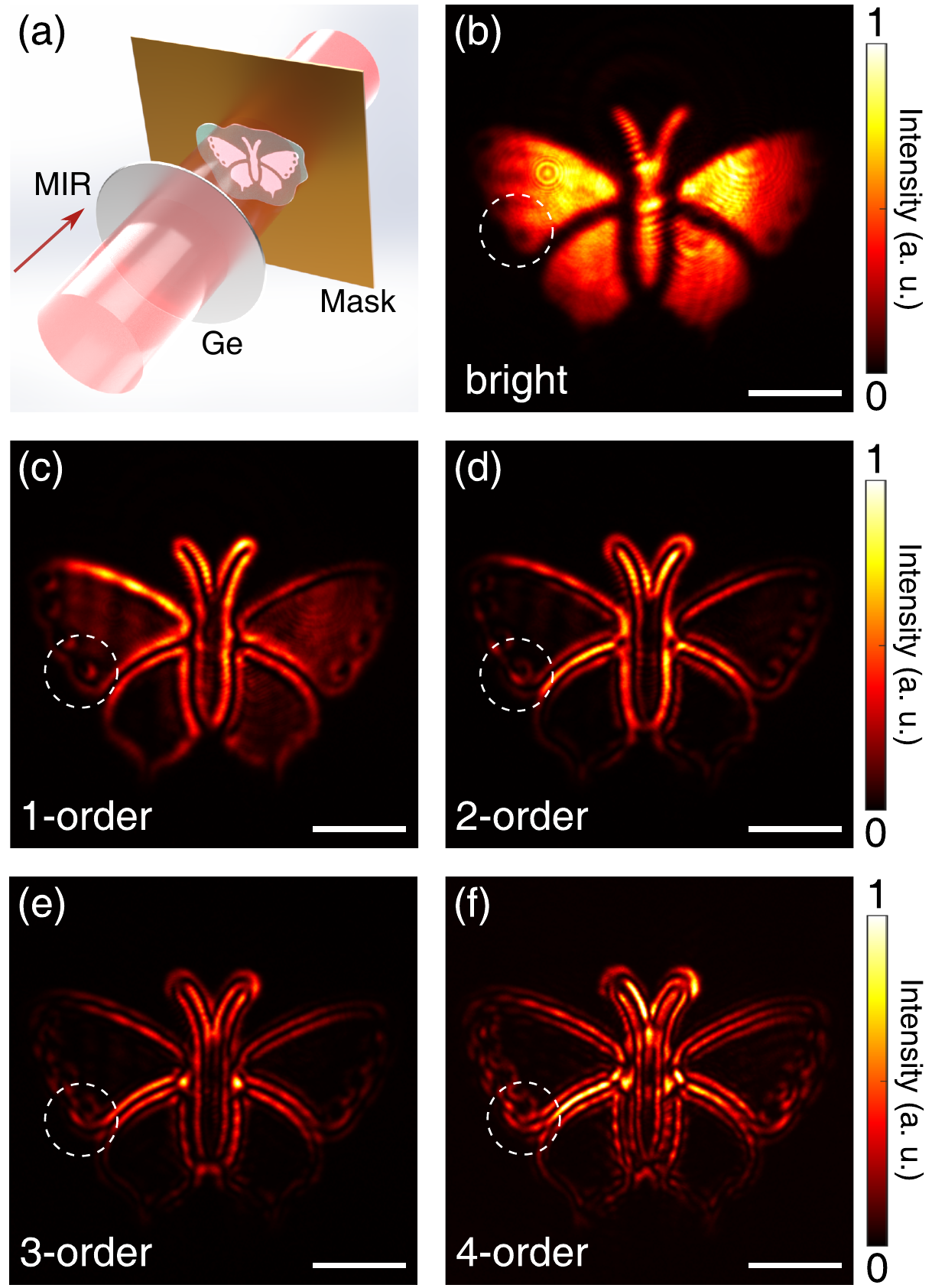}
	\caption{Performance of multiple-order MIR spatial differentiation for a transmission sample. (a) Schematic of the experimental configuration. A MIR beam passes through a 2 mm-thick germanium window and illuminates a butterfly-shaped mask coated with an irregular glue layer, which introduces additional amplitude and phase fluctuations. (b) Bright-field upconversion image showing blurred edges due to low-spatial-frequency background modulation induced by the glue. (c-f) First- to fourth-order isotropic differentiation images (scale bar: 1 mm). As the differentiation order increases, the glue-induced background is progressively suppressed. Consequently, edge details become increasingly pronounced, particularly at the wing tips as highlighted by dashed circles, demonstrating the superior edge-localization capability of higher-order spatial differentiation.}
	\label{fig5}
\end{figure}

\begin{figure}[t!]
	\centering
	\includegraphics[width=0.95\columnwidth]{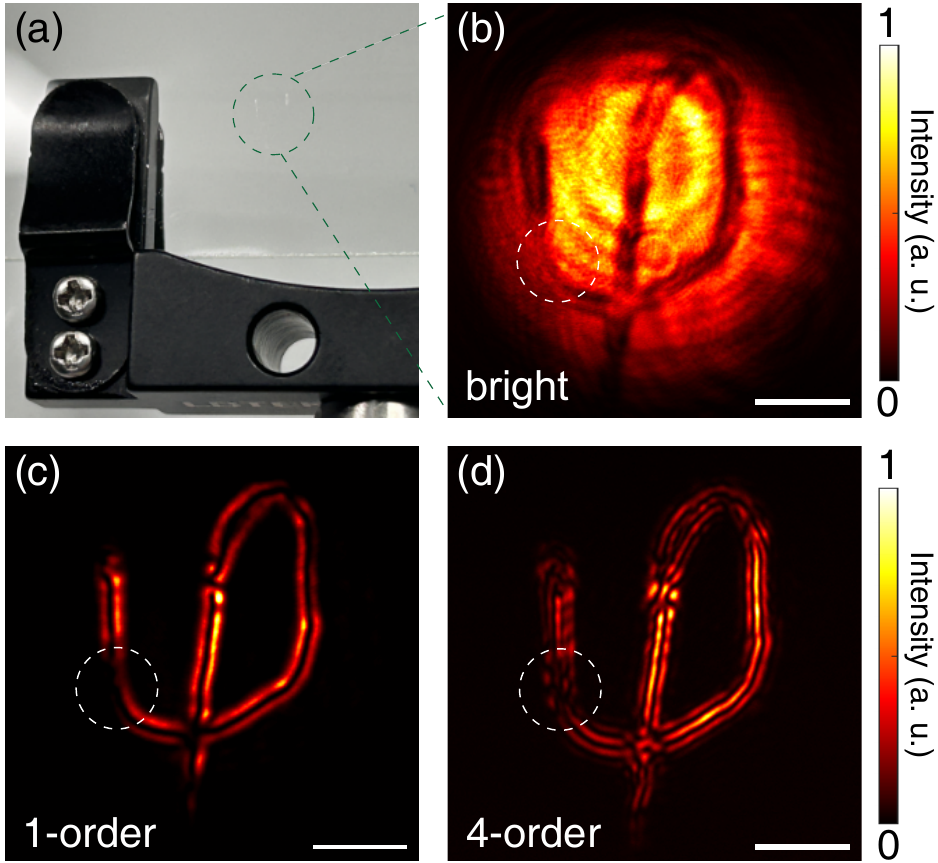}
	\caption{High-order MIR spatial differential imaging of a phase object. (a) A thin letter ``$\varphi$'' written in glue on a glass slide serves as a phase object that is nearly invisible under visible illumination. (b) Bright-field upconversion image showing only a faint outline due to the glue's weak MIR absorption; fine structural details are obscured by low-spatial-frequency background components. (c) First-order isotropic differentiation partially suppresses the background and improves edge contrast, though subtle features remain unresolved. (d) Fourth-order differentiation further removes residual low-frequency content and enhances high-frequency features, sharply revealing edges. The dashed circle highlights details that become clearly visible only at higher orders. The scale bar in each image corresponds to 1 mm.}
	\label{fig6}
\end{figure}

\subsection{Multiple-order MIR spatial differentiation for visualizing phase objects}
Finally, to further highlight the unique capabilities of high-order MIR differential imaging, we demonstrate its ability to penetrate obscuring layers, enhance contrast in nominally transparent samples, and sharpen edge features by suppressing the low-frequency background. As illustrated in Fig. \ref{fig5}(a), a butterfly-shaped transparent mask, coated with a UV-curable polymer whose thickness gradually increases from center to periphery, is placed behind a germanium plate that transmits only wavelengths above 2 $\mu$m. The composite object is illuminated with an MIR beam and visualized through the upconversion differential imaging system.

Figure \ref{fig5}(b) shows the corresponding bright-field image, where low-frequency intensity and phase fluctuations induced by the polymer layer degrade edge sharpness and reduce contrast. In contrast, Figs. \ref{fig5}(c-f) present the first- to fourth-order isotropic differential images. As the differentiation order increases, background interference from the UV layer is progressively suppressed, while high-spatial-frequency edge signals are enhanced. Particularly, filamentary structures near the butterfly wing tips become increasingly prominent at higher orders, which are barely discernible in the lower-order images. These results show that MIR high-order differential imaging significantly enhances edge clarity and feature visibility under complex, semi-transparent occlusion.

Figure \ref{fig6} further illustrates the system's ability to visualize pure phase objects and resolve subtle structural variations. A very thin layer of UV-curable glue, forming the letter ``$\varphi$'', is deposited on a transparent substrate. Due to the glue's near-transparency in the visible range, its pattern remains imperceptible under conventional illumination. However, MIR illumination reveals the object owing to its specific absorption in this spectral range. The bright-field upconversion image in Fig. \ref{fig6}(b) shows a faint outline of the object. When the first-order isotropic differential operator is applied, edge contrast improves significantly, clearly delineating the object's contour, as shown in Fig. \ref{fig6}(c). With fourth-order differentiation in Fig. \ref{fig6}(d), high-spatial-frequency features are further amplified, rendering the previously obscure structural details clearly resolvable. This confirms that higher-order operation exhibits stronger sensitivity to subtle phase or absorption gradients, enabling weak or slowly varying edge transitions to be clearly visualized. A more quantitative analysis is provided in Supplementary Note 6.

Such high-order spatial filtering is particularly advantageous for penetration imaging and subsurface defect detection in infrared-transparent materials like semiconductors and polymers. For example, germanium wafers are highly transparent above 2 $\mu$m, which is widely used in MIR photonic chip fabrication. Moreover, the inherent chemical specificity of MIR absorption lends itself to label-free imaging of biomedical specimens. The combination of molecular contrast with high-order edge enhancement may thus provide a powerful diagnostic tool for revealing subtle pathological features in complex biological environments.

\section{Conclusion}
Compared with existing infrared upconversion edge-enhanced imaging techniques, such as spiral phase contrast \cite{Wang2021LPR, Liu2019PRAppl, Qiu2018Optica, Gao2025OE}, amplitude-only Laplacian filters \cite{Junaid2020AO}, or angular spectrum selection \cite{Wei2024OL}, the proposed upconversion spatial differentiator achieves a substantial advancement in both flexibility and performance. By shaping the pump beam into a programmable complex-amplitude filter with the form $k_r^{\ell}e^{i\ell\phi}$, our system enables isotropic spatial differentiation of arbitrary order, surpassing previous methods that are typically limited to first-order or anisotropic edge detection. Furthermore, the upconversion architecture offers superior imaging performances by leveraging silicon-based cameras with high detection sensitivity and fast frame rate \cite{Huang2022NC, Dam2012NP, Zheng2024Optica}, outperforming conventional MIR focal plane arrays in dynamic range, noise characteristics, and temperature stability.

Building upon this foundation, several directions can further enhance the system's performance. First, introducing a 4f relay system to project the SLM phase directly onto the nonlinear interaction plane would minimize vortex beam divergence and significantly improve the usable spatial bandwidth and field of view \cite{Liu2019PRAppl}. Note that although a phase-only SLM is used, complex-amplitude modulation required for LG-like differentiators is realized through off-axis holographic encoding. Second, leveraging Fourier ptychography can synthetically enlarge the effective Fourier aperture, thereby overcoming the spatial-frequency truncation imposed by finite crystal thickness \cite{Zheng2024Optica}. This is particularly beneficial for high-order differentiation, where fine edge details correspond to higher spatial frequencies that are otherwise suppressed. Third, combining multiple poling periods using chirped or fan-out periodically poled lithium niobate crystals can expand the operational bandwidth and spatial acceptance angle \cite{Huang2022NC}. Additionally, integrating spectrally tunable MIR sources or supercontinuum generation would unlock the potential for wavelength-resolved higher-order differentiation \cite{Junaid2019Optica, Fang2024NC, Knez2022SA, Zhao2023NC}, paving the way for simultaneous spatial-spectral edge extraction in advanced imaging applications.

In summary, we have developed and experimentally demonstrated a high-order MIR upconversion imaging scheme based on topological complex-amplitude pump modulation. This system realizes reconfigurable, isotropic, and high-sensitivity spatial differentiation up to the fourth order, verified through both amplitude and phase imaging experiments. The method provides a versatile and scalable platform for MIR computational imaging and optical analog computing. Looking ahead, the synergistic advantages of broadband spectral coverage, precise edge discrimination, and high-speed tunability position the multiple-order MIR differentiating technique as a powerful tool to reveal fine structural features through obscuring layers or within transparent media, which holds particular promise for high-resolution, label-free analysis in both industrial and biomedical fields.

\section*{Acknowledgements}
This work was funded by Shanghai Pilot Program for Basic Research (TQ20220104); National Natural Science Foundation of China (62175064, 62235019, 62035005); Innovation Program for Quantum Science and Technology (2023ZD0301000); Shanghai Municipal Science and Technology Major Project (2019SHZDZX01); Natural Science Foundation of Chongqing (CSTB2025NSCQ-GPX0443); Postdoctoral Fellowship Program and China Postdoctoral Science Foundation (GZC20250545, 2024M760918, 2025T180224); Fundamental Research Funds for the Central Universities.

\section*{Conflict of Interest}
The authors declare no conflict of interests.

\section*{Supporting Information}
Supporting Information is accompanied to present more theoretical and experimental details. 

\section*{Data Availability Statement}
The data that support the findings of this study are available from the corresponding author upon reasonable request.

\section*{Keywords}
edge-enhanced detection, mid-infrared imaging, spatial differentiator, frequency upconversion

\end{document}